\documentclass[english]{revtex4}
\usepackage[T1]{fontenc}
\usepackage[latin9]{inputenc}
\usepackage{amssymb}
\usepackage{babel}

\begin{document}

\title{Temperature for the $\left(2+1\right)$-dimensional Black Hole with
Non Linear Electrodynamics from the Generalized Uncertainty Principle}

\author{Alexis Larrañaga}

\affiliation{Universidad Nacional de Colombia. Observatorio Astronómico Nacional
(OAN)}

\affiliation{Universidad Distrital Francisco José de Caldas. Facultad de Ingeniería.
Proyecto Curricular de Ingeniería Electrónica}

\email{ealarranaga@unal.edu.co}

\author{Hector J. Hortua}

\affiliation{Universidad Nacional de Colombia. Observatorio Astronómico Nacional
(OAN) }

\email{hjhortuao@unal.edu.co}

\begin{abstract}
In this paper, we study the thermodynamical properties of the $\left(2+1\right)$
dimensional black hole with a non-linear electrodynamics with a negative
cosmological constant, using the Generalized Uncertainty Principle
(GUP). This approach shows that there is a minimum mass or remnant
for the black hole, corresponding to the minimum radius of the event
horizon that has a size of the order of the Planck scale. We also show that the heat
capacity for this black hole is always positive.
\end{abstract}
\maketitle

Working in $\left(2+1\right)$ dimensional gravity, and using as the
source of the Einstein equations the stress-energy tensor of non-linear
electrodynamics, Cataldo et. al. \cite{cataldo} found a solution
with a Coulomb-like electric field (proportional to the inverse of
$r^{2}$), that describes charged-AdS space when considering a negative
cosmological constant. 

As is well known, the thermodynamical properties of black holes are
associated with the presence of the event horizon. Recently \cite{larranagagarcia},
it has been shown that the $\left(2+1\right)$ dimensional field equations
for the black hole with nonlinear electrodynamics can be interpreted
as the differential first law with the usual form 

\begin{equation}
dM=TdS+\Phi dQ,\end{equation}
where $T$ is the Hawking temperature that can be expressed in terms
of the surface gravity at the horizon $\kappa$ by

\begin{equation}
T=\frac{\kappa}{2\pi}.\label{eq:hawkingtemp}\end{equation}

On the other hand, in recent years the uncertainty relation that includes
gravity effects, known as the Generalized Uncertainty Principle (GUP),
has shown interesting results in the context of black hole evaporation
\cite{arraut}, extending the relation between temperature and mass
to scales of the order of the Planck lenght, $l_{p}=1.61\times10^{-33}cm$.
This treatment imply that $l_{p}$ is the smallest length scale in
the theory and it is related to the existence of an extreme mass (the
Planck mass $m_{p}=1.22\times10^{19}GeV$, which becomes the black
hole remnant), that corresponds to the maximum possible temperature.
This approach has been used recently \cite{larranaga} to extend
the temperature-masss relation for the $\left(2+1\right)$ dimensional
black hole with a nonlinear electric field and without cosmological
constant $\left(\Lambda=0\right)$. There is also shown that there
is a maximum temperature permited for the black hole that corresponds
to a horizon with size in the Planck scale.

In this paper, we investigate the thermodynamics of the $\left(2+1\right)$-dimensional
black hole with a nonlinear electric field reported in \cite{cataldo},
to show how the GUP extended to gravity in spaces with cosmological
constant as proposed by Arraut et. al. \cite{arraut} can be used
to calculate the Hawking temperature associated with the black hole.
The $T\left(M\right)$ equation gives the usual relation for large
masses but gets deformed when the mass becomes small. We also show
how there is a minimum mass for the black hole that corresponds to
a horizon with size in the Planck scale. Finally we calculate the
heat capacity for this black hole, to show that it is always positive.

\section{The 3-dimensional Black Hole with non-linear Electrodynamics}

The metric reported by Cataldo et. al. \cite{cataldo} is a solution
of the $\left(2+1\right)$ dimensional Einstein's field equations
with a negative cosmological constant $\Lambda<0$,

\begin{equation}
G_{\mu\nu}+\Lambda g_{\mu\nu}=8\pi T_{\mu\nu},\end{equation}
where we have used units such that $G=1$. To obtain a Coulomb-like
electric field, Cataldo et. al. used a nonlinear electodynamics, in
which the electromagnetic action does not depend only on the invariant
$F=\frac{1}{4}F_{\mu\nu}F^{\mu\nu}$, but it is a function of $F^{3/4}$;
and the energy-momentum tensor is restricted to be traceless. The
static circularly symmetric solution obtained has the line element 

\begin{equation}
ds^{2}=-f\left(r\right)dt^{2}+\frac{dr^{2}}{f\left(r\right)}+r^{2}d\varphi^{2},\label{eq:staticmetric}\end{equation}
where

\begin{equation}
f\left(r\right)=-M-\Lambda r^{2}+\frac{4Q^{2}}{3r}.\end{equation}

The electric field for this solution is

\begin{equation}
E\left(r\right)=\frac{Q}{r^{2}},\label{eq:electric field}\end{equation}

which is the standard Coulomb field for a point charge. Note that
the metric depends on two parameters $Q$ and $M$, that are identified
as the electric charge and the mass, respectively.

\subsection{Horizons}

The horizons of this solution are defiened by the condition \begin{equation}
f\left(r\right)=0\end{equation}
 or

\begin{equation}
-M-\Lambda r^{2}+\frac{4Q^{2}}{3r}=0.\label{eq:horizonmass}\end{equation}
For a cosmological constant in the range

\begin{equation}
-\frac{M^{3}}{12Q^{4}}\leq\Lambda<0,\end{equation}
the roots of the third-order polynomial (\ref{eq:horizonmass}) are
all real, and can be written as \cite{larranagagarcia}

\begin{eqnarray}
r_{1} & = & -2\sqrt{-\frac{M}{3\Lambda}}\cos\left(\frac{1}{3}\cos^{-1}\left(2\frac{Q^{2}}{M}\sqrt{-\frac{3\Lambda}{M}}\right)\right)\label{eq:r1}\\
r_{2} & = & -2\sqrt{-\frac{M}{3\Lambda}}\cos\left(\frac{1}{3}\cos^{-1}\left(2\frac{Q^{2}}{M}\sqrt{-\frac{3\Lambda}{M}}\right)+\frac{2\pi}{3}\right)\label{eq:r2}\\
r_{3} & = & -2\sqrt{-\frac{M}{3\Lambda}}\cos\left(\frac{1}{3}\cos^{-1}\left(2\frac{Q^{2}}{M}\sqrt{-\frac{3\Lambda}{M}}\right)+\frac{4\pi}{3}\right).\label{eq:r3}\end{eqnarray}

Note that the form of this solution impose the condition

\begin{equation}
2\frac{Q^{2}}{M}\sqrt{-\frac{3\Lambda}{M}}\leq1.\end{equation}
The equal sign defines a extreme black hole, with the maximum mass

\begin{equation}
M^{max}=\sqrt[3]{-12Q^{4}\Lambda}\end{equation}
and with the horizons

\begin{eqnarray}
r_{1} & = & -2\sqrt{-\frac{M^{max}}{3\Lambda}}\\
r_{2} & = & r_{3}=2\sqrt{-\frac{M^{max}}{3\Lambda}}.\end{eqnarray}
Note that the radius $r_{1}$ is negative, so it does not represent
a physical horizon because for other values of the mass it is always
negative. Therefore, this kind of black holes have only two physical
horizons, $r_{2}$ and $r_{3}$, that, for the extremal black hole,
coincide,

\begin{equation}
r_{2}=r_{3}=r_{ext}=-\left(\frac{2Q^{2}}{3\Lambda}\right)^{1/3}.\end{equation}
In the general case, the largest radius between $r_{2}$ and $r_{3}$
corresponds to the event horizon of the black hole $r_{+}$, while
the other corresponds to the inner horizon $r_{-}$.

Using the expansion $\cos^{-1}x\approx\frac{\pi}{2}-x-\frac{1}{6}x^{3}$,
the radii of the horizons $r_{2}$ and $r_{3}$ in equations (\ref{eq:r2})
and (\ref{eq:r3}) can be approximated up to the third order in

\begin{equation}
x=2\frac{Q^{2}}{M}\sqrt{-\frac{3\Lambda}{M}},\end{equation}
as

\begin{eqnarray}
r_{2}=r_{+} & \approx & \sqrt{-\frac{M}{3\Lambda}}\left[\sqrt{3}-\frac{1}{3}x+\frac{\sqrt{3}}{18}x^{2}+\frac{4}{81}x^{3}\right]+O\left(x^{4}\right)\label{eq:r2app}\\
r_{3}=r_{-} & \approx & \sqrt{-\frac{M}{3\Lambda}}\left[\frac{2}{3}x+\frac{8}{81}x^{3}\right]+O\left(x^{4}\right).\label{eq:r3app}\end{eqnarray}

If we consider only the first contribution in $x$ we get 

\begin{eqnarray}
r_{2}=r_{+} & \approx & \sqrt{-\frac{M}{\Lambda}}-\frac{2}{3}\frac{Q^{2}}{M}\label{eq:r2app2}\\
r_{3}=r_{-} & \approx & \frac{4}{3}\frac{Q^{2}}{M}.\label{eq:r3app2}\end{eqnarray}

\section{Hawking Temperature}

In this section we will deduce the Hawking temperature for this black
hole by using the usual definition (\ref{eq:hawkingtemp}). The surface
gravity can be calculated as

\begin{equation}
\kappa=\chi\left(x^{\mu}\right)a,\end{equation}
where $a$ is the magnitude of the four-acceleration and $\chi$ is
the red-shift factor. In order to calculate $\chi$, we will consider
a static observer, for whom the red-shift factor is just the proportionality
factor between the timelike Killing vector $K^{\mu}$ and the four-velocity
$V^{\mu}$, i.e.

\begin{equation}
K^{\mu}=\chi V^{\mu}.\end{equation}
 The metric (\ref{eq:staticmetric}) has the Killing vector

\begin{equation}
K^{\mu}=\left(1,0,0\right)\end{equation}
while the four-velocity is calculated as

\begin{equation}
V^{\mu}=\frac{dx^{\mu}}{d\tau}=\left(\frac{dt}{d\tau},0,0\right).\end{equation}
This gives

\begin{equation}
V^{\mu}=\left(f^{-1}\left(r\right),0,0\right)=\left(\frac{1}{\sqrt{-M-\Lambda r^{2}+\frac{4Q^{2}}{3r}}},0,0\right),\end{equation}
and therefore, the red-shift factor is

\begin{equation}
\chi\left(r\right)=\sqrt{-M-\Lambda r^{2}+\frac{4Q^{2}}{3r}}.\end{equation}
 On the other hand, the four-acceleration is given by

\begin{equation}
a^{\mu}=\frac{dV^{\mu}}{d\tau},\end{equation}
that has components

\begin{eqnarray}
a^{0}=a^{\varphi} & = & 0\\
a^{r} & = & -\Lambda r-\frac{2}{3}\frac{Q^{2}}{r^{2}}.\end{eqnarray}
and therefore, the magnitude of the four-acceleration is

\begin{equation}
a=\sqrt{g_{\mu\nu}a^{\mu}a^{\nu}}=\frac{-\Lambda r-\frac{2}{3}\frac{Q^{2}}{r^{2}}}{\sqrt{-M-\Lambda r^{2}+\frac{4Q^{2}}{3r}}}.\end{equation}

Then, the surface gravity at the event horizon is given by the absolute
value

\begin{equation}
\kappa_{+}=\left|-\Lambda r-\frac{2}{3}\frac{Q^{2}}{r^{2}}\right|_{r=r_{+}}.\end{equation}

Note that for $r=r_{ext}$, the surface gravity is $\kappa\left(r_{max}\right)=0$,
i.e. that the extreme black hole has zero Hawking temperature,

\begin{equation}
T\left(M^{max}\right)=0.\end{equation}

Using the approximate forms for $r_{+}$ and $r_{-}$ given by equations
(\ref{eq:r2app}) and (\ref{eq:r3app}), we can write the Hawking
temperature as a function of the mass. For the event horizon we have,
up to second order in $x$,

\begin{equation}
T_{+}\left(M\right)=\frac{\kappa_{+}}{2\pi}=\frac{1}{2\pi}\left|-\Lambda r_{+}-\frac{2}{3}\frac{Q^{2}}{r_{+}^{2}}\right|\end{equation}

\begin{equation}
T_{+}\left(M\right)\approx\frac{1}{2\pi}\left|-\Lambda\sqrt{-\frac{M}{3\Lambda}}\left[\sqrt{3}-\frac{1}{3}x\right]-\frac{2}{3}Q^{2}\left(-\frac{M}{3\Lambda}\right)^{-1}\left[\sqrt{3}-\frac{1}{3}x\right]^{-2}\right|\end{equation}

\begin{equation}
T_{+}\left(M\right)\approx\frac{1}{2\pi}\left|\sqrt{-M\Lambda}-\sqrt{-\frac{M\Lambda}{3}}\frac{1}{3}x+\frac{2\Lambda Q^{2}}{M}\left(3-\frac{2x}{\sqrt{3}}\right)^{-1}\right|\end{equation}

\begin{equation}
T_{+}\left(M\right)\approx\frac{1}{2\pi}\left|\sqrt{-M\Lambda}-\sqrt{-\frac{M\Lambda}{3}}\frac{1}{3}x+\frac{2\sqrt{3}\Lambda Q^{2}}{M\left(3\sqrt{3}-2x\right)}\right|\end{equation}

\begin{equation}
T_{+}\left(M\right)\approx\frac{1}{2\pi}\left|\sqrt{-M\Lambda}-\frac{2}{3}\frac{\Lambda Q^{2}}{M}+\frac{2\sqrt{-M\Lambda}Q^{2}}{3\sqrt{-\frac{M^{3}}{\Lambda}}-4Q^{2}}\right|.\label{eq:bhtemperature}\end{equation}
Note that, for $Q=0$, this expression becomes

\begin{equation}
T_{+}^{BTZ}\left(M\right)=\frac{\sqrt{-M\Lambda}}{2\pi},\end{equation}
that corresponds to the temperature at the horizon of the static non-charged
BTZ black hole \cite{akbar}. On the other hand, for the inner horizon
$r_{-}$ we have\begin{equation}
T_{-}\left(M\right)=\frac{\kappa_{-}}{2\pi}=\frac{1}{2\pi}\left|-\Lambda r_{-}-\frac{2}{3}\frac{Q^{2}}{r_{-}^{2}}\right|\end{equation}

\begin{equation}
T_{-}\left(M\right)\approx\frac{1}{2\pi}\left|-\Lambda\sqrt{-\frac{M}{3\Lambda}}\frac{2}{3}x-\frac{2}{3}Q^{2}\left[\sqrt{-\frac{M}{3\Lambda}}\frac{2}{3}x\right]^{-2}\right|\end{equation}

\begin{equation}
T_{-}\left(M\right)\approx\frac{1}{2\pi}\left|-\Lambda\sqrt{-\frac{M}{3\Lambda}}\frac{2}{3}x+\frac{9\Lambda Q^{2}}{2Mx^{2}}\right|\end{equation}

\begin{equation}
T_{-}\left(M\right)\approx\frac{1}{2\pi}\left|-\frac{4}{3}\frac{\Lambda Q^{2}}{M}-\frac{3M^{2}}{8Q^{2}}\right|.\end{equation}

\section{Hawking Radiation and the Generalized Uncertainty Principle}

Now we will consider the GUP and the Hawking radiation derived from
it. We will show how the GUP will produce a deformation in the temperature-mass
relation when considered close to the Planck lenght. As stated by
Arraut et. al. \cite{arraut} the GUP can be stated, in the case
of a non-zero cosmological constant and in units with $c=\hbar=1$,
by the relation

\begin{equation}
\Delta x\Delta p\gtrsim\frac{1}{2}+\frac{G}{2}\left(\Delta p\right)^{2}-\frac{\gamma}{3\Lambda}\frac{1}{\left(\Delta p\right)^{2}},\end{equation}
where $G$ is the gravitational constant, $\Lambda$ is cosmological
constant and $\gamma$ is a constant factor which accounts for the
fact that the GUP equation is dealing with orders of magnitudes estimates.
When comparing the GUP results with the above Hawking temperature,
we will see that $\gamma$ is expected to be of the order of 1. Since
the Planck lenght $l_{p}$ can be written as

\begin{equation}
l_{p}^{2}=G\hbar=\frac{\hbar^{2}}{m_{p}^{2}},\end{equation}
where $m_{p}$ is the mass of Planck, the GUP can be written as

\begin{equation}
\Delta x\Delta p\gtrsim\frac{1}{2}+\frac{l_{p}^{2}}{2}\left(\Delta p\right)^{2}-\frac{\gamma}{3\Lambda}\frac{1}{\left(\Delta p\right)^{2}},\end{equation}

or

\begin{equation}
\Delta x\Delta p\gtrsim\frac{1}{2}+\frac{1}{2}\frac{\left(\Delta p\right)^{2}}{m_{p}^{2}}-\frac{\gamma}{3\Lambda}\frac{1}{\left(\Delta p\right)^{2}}.\end{equation}

To apply this uncertainty principle to the black hole evaporation
process consist in identifying the $\Delta x$ with the event horizon
radius $r_{+}$ and the momentum $\Delta p$ with the Hawking temperature
up to a $2\pi$ factor \cite{myung}$\left(\Delta P\sim2\pi T\right)$.
Therefore, we can write the GUP as a fourth order equation for the
temperature,

\begin{equation}
r_{+}2\pi T=\frac{1}{2}+\frac{2\pi^{2}T^{2}}{m_{p}^{2}}-\frac{\gamma}{3\Lambda}\frac{1}{4\pi^{2}T^{2}}\end{equation}

For high temperatures, the GUP equation can be approximated by

\begin{equation}
r_{+}2\pi T=\frac{1}{2}+\frac{2\pi^{2}T^{2}}{m_{p}^{2}}\end{equation}
that corresponds to a quadratic polynomial for $T$,

\begin{equation}
\frac{4\pi^{2}}{m_{p}^{2}}T^{2}-4\pi r_{+}T+1=0\end{equation}

From which it follows that the temperature-mass relation is

\begin{equation}
T\left(M\right)=\frac{m_{p}^{2}r_{+}}{2\pi}\left[1\pm\sqrt{r_{+}^{2}-\frac{1}{m_{p}^{2}}}\right].\label{eq:bhtemperature2}\end{equation}
Note that the argument in the square root defines a minimum radius
for the event horizon,

\begin{equation}
r_{+}^{min}=\frac{1}{m_{p}}=l_{p},\label{eq:rmin}\end{equation}
i.e. that $r_{+}$ can not be smaller than the Planck lenght. This
lenght corresponds to a minimum mass or remnant for the black hole
of the order of

\begin{equation}
M^{min}=\Lambda l_{p}^{2}.\end{equation}
Using equation (\ref{eq:bhtemperature}), the emperature of the black
hole corresponding to this mass is

\begin{equation}
T_{+}\left(M^{min}\right)\approx\frac{1}{2\pi}\left|\left|\Lambda\right|l_{p}-\frac{2}{3}\frac{Q^{2}}{l_{p}^{2}}+\frac{2\left|\Lambda\right|l_{p}Q^{2}}{3\left|\Lambda\right|l_{p}^{3}-4Q^{2}}\right|.\end{equation}
Note that this expression gives, in the $Q=0$ case, a minimum mass
for the BTZ black hole,

\begin{equation}
T_{+}^{BTZ}\left(M^{min}\right)=\frac{\left|\Lambda\right|l_{p}}{2\pi}.\end{equation}

Finally, the heat capacity for the $\Lambda=0$ black hole can be
calculated using equation (\ref{eq:bhtemperature2}), 

\begin{eqnarray}
C_{Q} & = & \left(\frac{\partial M}{\partial T}\right)_{Q}\\
 & = & \frac{1}{2\pi m_{p}^{2}}\frac{\sqrt{r_{+}^{2}-\frac{1}{m_{p}^{2}}}}{2r_{+}^{2}-\frac{1}{m_{p}^{2}}}\left(-\frac{1}{2\Lambda}\sqrt{-\frac{\Lambda}{M}}+\frac{2}{3}\frac{Q^{2}}{M^{2}}\right)^{-1}.\end{eqnarray}
Note that the heat capacity of this black hole is negative, $\frac{\partial T}{\partial M}<0$,
when

\begin{equation}
r_{+}^{2}<\frac{1}{2m_{p}^{2}}=\frac{l_{p}^{2}}{2},\end{equation}
but condition (\ref{eq:rmin}) impose that the minimum radius is $r_{+}^{min}=l_{p}$.
Therefore the heat capacity of this black hole is always positive.

\section{Conclusion}

We have studied the thermodynamics of the $\left(2+1\right)$ dimensional
black hole with non-linear electrodynamics and with a negative cosmological
constant using the Generalized Uncertainty Principle. This gives a
minimum mass or remnant for the black hole, that corresponds to a
maximum Hawking temperature depending only on the electric charge
$Q$. The solution with the maximum temperature is obtained when the
black hole has a size of the order of the Planck scale (minimum horizon). 

Equation (\ref{eq:bhtemperature2}) gives the temperature-mass relation,
and as is shown, it gives the standard Hawking temperature (\ref{eq:bhtemperature})
for small masses, but gets deformed for masses close to $M^{min}$.
Finally, the heat capacity of this black hole is always positive.
This analysis confirms that Planck lenght seems to be the smallest
lenght in nature, even in $\left(2+1\right)$ dimensions.

\end{document}